\documentclass[usenatbib, usegraphicx]{mn2e}

\begin{document}

\title{Tracing sunspot groups to determine angular momentum transfer on the Sun}
\author[D. Sudar et al.]{D. Sudar$^{1}$
        I. Skoki\'{c}$^{2}$
        D. Ru\v{z}djak$^{1}$
        R. Braj\v{s}a$^{1}$
        H. W\"{o}hl$^{3}$
\\
$^1$ Hvar Observatory, Faculty of Geodesy,
              Ka\v{c}i\'{c}eva 26, University of Zagreb, 10000 Zagreb, Croatia\\
$^2$ Cybrotech Ltd, Bohinjska 11, 10000 Zagreb, Croatia\\
$^3$ Kiepenheuer-Institut f\"{u}r Sonnenphysik,
             Sch\"{o}neckstr. 6, 79104 Freiburg, Germany
}

   \date{Release \today}
   \maketitle

\begin{abstract}
The goal of this paper is to investigate Reynolds stresses and to check if it
is plausible that they are responsible for angular momentum transfer toward the
solar equator. We also analysed meridional velocity, rotation velocity residuals
and correlation between the velocities.
We used sunspot groups position measurements from GPR (Greenwich Photographic Result) and 
SOON/USAF/NOAA (Solar Observing Optical Network/United States Air Force/National
Oceanic and Atmospheric Administration) databases covering the period from 1878 until 2011.
In order to calculate velocities we used daily motion of sunspot groups. The sample was also limited to
$\pm$58\degr in Central Meridian Distance (CMD) in order to avoid solar limb effects. We mainly investigated velocity patterns
depending on solar cycle phase and latitude.
We found that meridional motion of sunspot groups is toward the centre of activity from
all available latitudes and in all phases of the solar cycle.
The range of meridional velocities is $\pm10$ m s$^{-1}$.
Horizontal Reynolds stress is negative
at all available latitudes and indicates that there is a 
minimum value ($q\approx$ - 3000 m$^2$ s$^{-2}$) located at $b\approx\pm$30$\degr$.
In our convention this means that angular
momentum is transported toward the solar equator in agreement with the observed rotational profile
of the Sun.
\end{abstract}

\section{Introduction}
Tracing the motion of sunspot groups has a long history and is still used frequently today
for studies of solar rotation and related phenomena.
In this work we used daily motion of sunspot groups from GPR (Greenwich Photographic Result)
and SOON/USAF/NOAA (Solar Observing Optical Network/United States Air Force/National
Oceanic and Atmospheric Administration) databases combined into a single dataset.
The key parameter we will investigate is horizontal Reynolds stress which might
explain the transfer of angular momentum toward the equator.
In a review by \citet{Schroeter1985}, the author advocated the use of sunspot data since both components
of horizontal Reynolds stress can be measured separately.
GPR dataset complemented by SOON/USAF/NOAA is the longest homogeneous sunspot catalogue
and presents a unique opportunity to study long term average properties of the solar velocity
field as well as its variations on time scale of a century.
The GPR dataset we used was digitised in the frame of several projects and it is not identical
to the on-line version\footnote{http://solarscience.msfc.nasa.gov/greenwch.shtml}.
Various analyses were carried out by using whole or parts of our GPR dataset
(\citealt{Balthasar1980, Balthasar1981, Arevalo1982, Arevalo1983}; \citealt*{Balthasar1986a, Balthasar1987};
\citealt{Brajsa2000, Wohl2001, Brajsa2002, Brajsa2004, Ruzdjak2004, Ruzdjak2005, Brajsa2006, Brajsa2007}).

A very comprehensive analysis of cycle to cycle variations of rotation velocity for
GPR dataset is given in \citet{Balthasar1986a} and \citet{Brajsa2006}. The same authors analysed also rotation
velocity variations with respect to phase of the solar cycle.

Recently, a comprehensive project of revising the GPR dataset was undertaken
\citep{Willis2013a, Willis2013b, Erwin2013}
to correct a number of erroneous measurements and typos which illustrates the importance of
the GPR dataset.

Various aspects of solar rotation and related phenomena are used to quantify and constrain solar models.
Usually the main focus of such investigations are meridional motions and rotation
residual velocity. Correlation between two velocities and their covariance are
of even greater significance. All of these quantities play an important part
in understanding the solar cycle and its variations from one cycle to another.
Series of numerical simulations and theoretical works regarding the transfer of angular momentum toward the equator
has been carried out by many authors \citep*{Canuto1994, Chan2001, Rudiger2004, Kapyla2004, Hupfer2006}.

Studies of meridional flows show wide discrepancy in their results both qualitatively
and quantitatively.
By analysing sunspot groups data obtained at Mount Wilson \citet{Howard1986} found that for lower
solar latitudes ($b<\pm$15\degr) meridional
motion is negative, i.e. toward the equator. The magnitude of the effect being
$\omega_{mer}\approx$0.02\degr day$^{-1}$. Above mentioned latitude, the flow is toward the poles with
a possible hint that at even higher latitudes it changes sign again.
\citet{Wohl2001} studied meridional motion of stable recurrent sunspot
groups and they also found that the average flow is equatorward for lower latitudes, while
poleward motion occurs at latitudes higher than latitude of the centre of activity.
Moreover, the magnitude of the effect is the same as that of \citet{Howard1986}.
But their data does not provide any indication that the flow might change to equatorward
at even higher latitudes.
By studying sunspot drawings obtained at National Astronomical Observatory of Japan during the years 1954-1986,
\citet{Kambry1991} found that meridional flow is equatorward for latitudes in the range
-20\degr\ to +15\degr. They also found an indication of a solar-cycle dependence of meridional motions.
\citet{Howard1991a,Howard1996} found that sunspot groups move away from the average
latitude of activity while plages move toward it.

By using Doppler line shifts, \citet{Duvall1979} observed
roughly constant poleward flow on the order of $v_{mer}=$20 m s$^{-1}$ in the whole range of studied
latitudes (10\degr\ - 50\degr). \citet{Howard1979} mentions the same value, but referring
only  to {\sl higher latitudes}.
By analysing Doppler velocity data obtained with the Global Oscillating Network
Group (GONG) instruments, \citet{Hathaway1996} concluded that the flow
is poleward at all latitudes with typical values being about $v_{mer}=$20 m s$^{-1}$,
but with episodes of much stronger flows (60 m s$^{-1}$).
In contrast to the above,
\citet{Perez1981} found equatorward motion of $v_{mer}\approx$20 m s$^{-1}$.
\citet{Lustig1990} studied meridional plasma motions using Doppler line shifts
covering the period from 1982 until 1986, covering about a half of the solar
cycle. The authors concluded that systematic
meridional motion, if even present at all, can not be larger than $v_{mer}=$10 m s$^{-1}$ toward solar equator
for latitudes below $b=\pm$35\degr\ in both hemispheres.

Using high-resolution magnetograms taken from 1978 to 1990 with the NSO Vacuum Telescope on Kitt Peak,
\citet{Komm1993} observed poleward flow of the order of $v_{mer}=$10 m s$^{-1}$ in both hemispheres.
The flow increased in amplitude from 0 m s$^{-1}$ at the equator, reached a maximum at mid-latitude and
slowly decreased at even higher latitudes.
By applying time-distance helioseismology, \citet{Zhao2004} found poleward meridional flows
of the order of $v_{mer}=$20 m s$^{-1}$. In addition they found meridional circulation cells converging
toward the activity belts in both hemispheres.

Apart from evolutionary loss of angular momentum \citep[see, e.g.,][]{Guinan2009}, the Sun also exhibits changes of the rotational
profile on much smaller time scales.
A cyclic pattern with a period of $\approx$11 years of alternating faster and slower rotational bands is called
torsional oscillations.
Existence of torsional oscillations on the Sun were first reported by \citet{Howard1980}.
They were further confirmed by \citet{Ulrich1988}, \citet{Howe2000}, \citet{Haber2002},
\citet{Basu2003} and others.
At latitudes below $b\approx\pm$40\degr\ bands propagate equatorward while
each band is about 15\degr\ wide in latitude. The amplitude of the effect
is about $\Delta v_{rot}=$5-10 m s$^{-1}$.
\citet{Brajsa2006} showed an interesting analysis of rotation velocity residuals versus
phase of the solar cycle in their Fig.~6 showing variations of $\Delta\omega_{rot}\approx$0.05 $\degr$
day$^{-1}$ corresponding to $\Delta v_{rot}\approx$7 m s$^{-1}$ at the equator.

Transfer of the angular momentum from higher to lower latitudes
can be revealed by studying the correlation and covariance between azimuthal
and meridional flows. Covariance, denoted as $Q=<\Delta v_{rot}v_{mer}>$, is a
horizontal Reynolds stress.
Reynolds stress is thought to be the main generator of maintaining current
differential rotation profile \citep[see, e.g.,][]{Pulkkinen1998, Rudiger2004}.
Indeed, observations seems to show that the correct value of $Q$ was
observed \citep{Ward1965,Belvedere1976,Schroeter1976,Gilman1984,Pulkkinen1998,Vrsnak2003}. In
addition, some authors \citep{Ward1965,Gilman1984,Pulkkinen1998,Vrsnak2003}
investigated latitudinal dependence of Reynolds stress, and found that it mainly
decreases with higher latitudes with a possible minimum around $b=\pm30$\degr.

\section{Data and reduction methods}

We limited the data to $\pm$58\degr\ in Central Meridian Distance (CMD) which corresponds to about 0.85 of projected
solar radius \citep[cf.][]{Balthasar1986a}.
With such cutoff we obtained a sample of 92091 data pairs from GPR to obtain rotation rates
and meridional velocities.
We used two subsequent measurements of individual sunspot group to get one velocity value.

Using the same CMD cutoff, we ended up with a sample of 43583 data pairs from observations found in
SOON/USAF/NOAA database
in the period from 1977 until 2011.
Combining these samples into a single dataset
the total amount of data points for sunspot groups was 135674 spanning from year 1878 till 2011.
In the rest of this work we will refer to this combined dataset as EGR (Extended Greenwich Result).
In the GPR era (until 1977), positions of sunspot groups are given with accuracy of 0.1$\degr$ in both coordinates while
subsequent measurements were usually taken 1 day apart. After 1977 the positions are usually given
with the accuracy of 1.0$\degr$.

When observed by tracers, solar rotation and related phenomena should be treated
statistically which requires large number of measurements for proper analysis.
While solar rotation velocity has large signal to noise ratio, ($S/N$), solar rotation
residuals, meridional velocities and Reynolds stress are significantly weaker effects with lower
$S/N$. So, in this paper we will mostly concentrate to identify basic net effect in various
relationships between mentioned phenomena.

Meridional motion and angular rotation velocity were calculated from two subsequent
measurements of position. Since most of the measurements in our dataset are one day apart,
velocities are calculated from daily shifts of sunspot groups.
To obtain rotation velocity residuals it is necessary
to subtract the actual velocity measured from the average rotation velocity
at given latitude. Synodic angular velocities were calculated by using the daily motion
of sunspot groups and converted to sidereal angular velocities using the procedure
described in \citet{Rosa1995} and \citet{Brajsa2002}.
Due to latitudinal distribution of sunspots it is sufficient
to use only the first two terms in the standard solar differential rotation
equation:
\begin{equation}
\omega(b) = A + B\sin^{2}b,
\end{equation}
where $b$ is the heliographic latitude and $\omega(b)$ is sidereal angular velocity.

For the EGR dataset we obtained $A = 14.499\pm 0.005$\degr\ per day and $B = -2.64\pm0.05$\degr\ per day
that we calculated by fitting the above equation to all points in the dataset ($n=135674$).

After the subtraction was carried out angular velocity residuals have been transformed
to linear velocities residuals ($\Delta v_{rot}$) in meters per second,
taking into account latitudes of the tracers.
Solar radius used for conversion from angular velocities to linear
velocities was $R_{\odot}= 696.26\cdot10^{3}$ km \citep{Stix1989}.

We limited calculated sidereal rotation velocity to 8 - 19\degr\ per day in
order to eliminate any gross errors usually resulting from misidentification
of sunspot groups or typos.

Angular meridional velocities were also transformed to linear velocities in m s$^{-1}$.

\begin{figure}
\includegraphics[width=84mm]{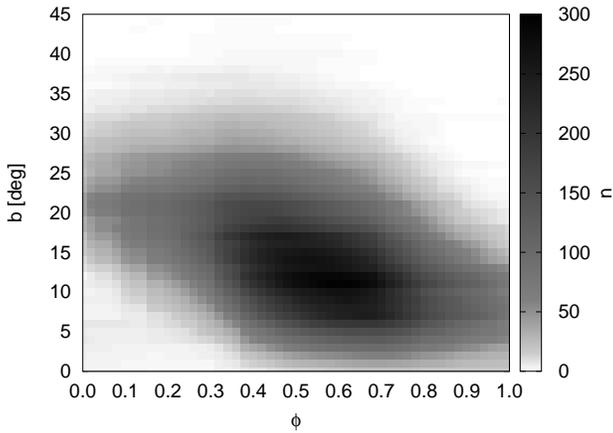}
\caption{Latitudinal distribution of sunspot groups from EGR dataset with respect to phase
of the solar cycle, $\phi$. All data are folded into one solar hemisphere.}
\label{Fig_butterfly}
\end{figure}
In the following section we will present several map plots of various quantities
depending on latitude, $b$, and phase of the solar cycle, $\phi$. Therefore, it is
useful to show latitudinal distribution of sunspots from EGR dataset with respect to phase
of the solar cycle, $\phi$, in order to indicate where the results are more reliable
(Fig.~\ref{Fig_butterfly}).
We folded all the data into one solar hemisphere.

In order to determine the phase we used times of minima
and maxima of solar activity found in Table 1 from
\citet{Brajsa2009}. All points that belong after the minimum of the solar cycle and before
maximum were mapped to $[0,0.5]$ phase range. Points after the maximum, but before
the minimum of the next cycle, were assigned phase in $[0.5,1]$ range.
Phase was calculated in a linear scale:
\begin{equation}
\phi_{i} = \frac{t_{i}-t_{min/max}}{t_{max/min} - t_{min/max}}.
\end{equation}

Since distribution of sunspots in latitude is not uniform, some precaution
must be used in order not to detect false motion \citep{Olemskoy2005}. 
Calculated velocities need to be assigned to some latitude. Considering
we have two measurements of position for one velocity, we have to decide
to which latitude we should assign the velocity. \citet{Olemskoy2005}
showed that false flow can arise if average latitude of
the two values is used, since the gradient of sunspot latitudinal distribution
will pollute the result. They also showed that this false meridional flow is of the right
order of magnitude and in the right direction as the results obtained by many authors using tracers to
detect surface flows on the sun. However, there is a simple solution to this problem:
if we assign the velocity to the latitude of the first measurement of position, there is no net flow
into the latitude bin from other latitudes and we don't have to worry about
the non-uniform distribution of sunspots in latitude. \citet{Olemskoy2005} also concluded the same.

\section{Results}
\subsection{Meridional flow}
We used the convention that negative meridional velocity reflects
equatorward motion: $v_{mer} = -\partial b/\partial t$ for southern
hemisphere, where we have defined southern latitudes as negative values.

\begin{figure}
\includegraphics[width=84mm]{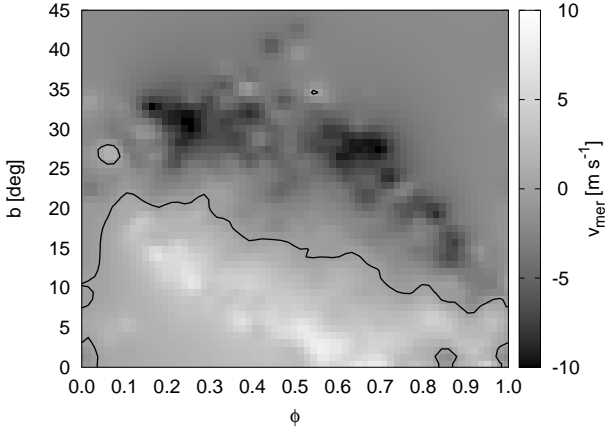}
\caption{Map plot of meridional velocity, $v_{mer}$, versus phase of the solar cycle, $\phi$,
and latitude, $b$. Also shown is a contour line where $v_{mer}=0$ with a solid line.}
\label{mapchart_vmer}
\end{figure}

In this section we investigate the properties of $v_{mer}$ depending on latitude, $b$,
and phase of the solar cycle, $\phi$. 
In Fig.~\ref{mapchart_vmer} we show a map plot of $v_{mer}$ versus cycle phase
and latitude, $b$, for EGR dataset. All points were folded into one phase diagram and both solar hemispheres
were folded together according to our convention above.
The map plot is constructed first by binning the data into square bins of width 0.1 in
phase of the cycle, $\phi$ and height 1\degr in latitude. Then we calculated the average values
of velocity in each bin, discarding all the bins where the number of data points was
less than 10 (see Fig.~\ref{Fig_butterfly}). Finally we calculated smoothed averages
of each bin with weight given by $w(d) = 1/(1 + d^{2})$,
where $d$ is a distance of each data point from the map grid point.
Brighter shades of grey depict polewards motion
($v_{mer} > 0$), while darker shades show motion toward the solar equator ($v_{mer} < 0$).
On the same plot we show a contour line of $v_{mer} = 0$ with a solid line which
clearly separates the two regions of opposite meridional flow.

\begin{figure}
\includegraphics[width=84mm]{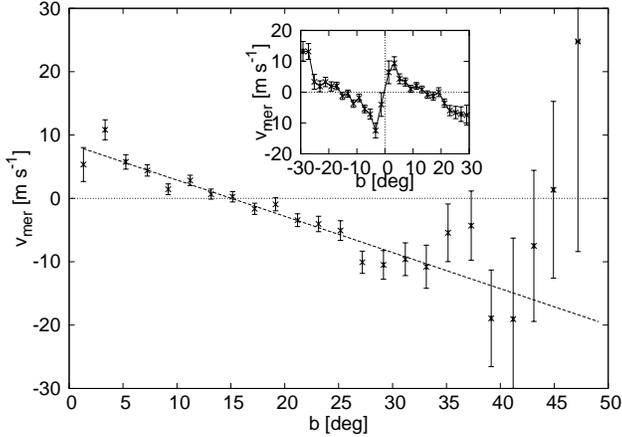}
\caption{Average values of $v_{mer}$ in bins 2\degr\ wide in latitude, $b$.
Linear fit (Eq.~\ref{vmer_b_lin_fit}) is shown with a dashed line. In the inset
we show average values $v_{mer}$ for North and South solar hemisphere separated.}
\label{vmer_b_aver}
\end{figure}

Such distinct appearance can be easily confirmed by plotting average values of $v_{mer}$
in bins 2\degr\ wide in latitude (Fig.~\ref{vmer_b_aver}). This is similar to
Fig.~\ref{mapchart_vmer} only integrated in phase of the solar cycle, $\phi$, or
in other words integrated in time.
In the same Fig. we show a linear fit of $v_{mer}(b)$ through individual data points, given by
equation:
\begin{equation}
v_{mer}=(-0.571\pm0.038)\ \mathrm{m\ s}^{-1}\mathrm{(\degr)}^{-1}\cdot b + (8.61\pm0.64)\ \mathrm{m\ s}^{-1}.
\label{vmer_b_lin_fit}
\end{equation}
From the above equation we get the intersect with
x-axis, $v_{mer}(b) = 0$, for $b\approx$ 15\degr. In the inset to Fig.~\ref{vmer_b_aver} we show
average values of $v_{mer}$ separately for the two solar hemispheres. We can see that for very low latitudes
$v_{mer}$ goes back to zero and reverses sign when crossing the solar equator.

We also divided the data into 10 bins in phase, $\phi$, each being 0.1 wide. Then we
calculated $v_{mer}(b)$ linear fits for those 10 bins.
The coefficients, given by equation:
\begin{equation}
v_{mer}=c_{1} b + c_{2},
\label{Eq_c1_c2}
\end{equation}
are shown in Table~\ref{linFitCoeffs}.

\begin{table}
\caption{Values of linear fit coefficients (Eq.~\ref{Eq_c1_c2}), intersect
with x-axis, $b_{v_{mer}=0}$, and average latitude of sunspot groups, $\overline{b(\phi)}$.}
\label{linFitCoeffs}
\begin{center}
\begin{tabular}{c c c c c}
\hline
$\phi$ & $c_{1}$[m s$^{-1}$ (\degr)$^{-1}$] & $c_{2}$[m s$^{-1}$] & $b_{v_{mer}=0}$[\degr] & $\overline{b(\phi)}[\degr]$\\
\hline
0.05 & -0.11$\pm$0.22 & 0.3$\pm$4.8 & 2.7$\pm$43.8 & 19.3\\
0.15 & -0.51$\pm$0.23 & 10.3$\pm$5.5 & 19.9$\pm$13.9 & 22.0\\
0.25 & -0.94$\pm$0.17 & 20.5$\pm$3.7 & 21.8$\pm$5.5 & 21.1\\
0.35 & -0.83$\pm$0.12 & 15.4$\pm$2.5 & 18.6$\pm$4.0 & 19.1\\
0.45 & -0.65$\pm$0.11 & 10.6$\pm$1.9 & 16.3$\pm$4.0 & 16.9\\
0.55 & -0.63$\pm$0.09 & 10.2$\pm$1.3 & 16.1$\pm$2.9 & 15.0\\
0.65 & -0.73$\pm$0.10 & 9.6$\pm$1.4 & 13.2$\pm$2.6 & 13.1\\
0.75 & -0.88$\pm$0.15 & 9.2$\pm$1.8 & 10.4$\pm$2.7 & 11.4\\
0.85 & -1.20$\pm$0.24 & 11.1$\pm$2.7 & 9.3$\pm$2.9 & 9.9\\
0.95 & -0.73$\pm$0.30 & 5.8$\pm$3.4 & 8.0$\pm$5.6 & 9.5\\
\hline
\end{tabular}
\end{center}
\end{table}

\begin{figure}
\includegraphics[width=84mm]{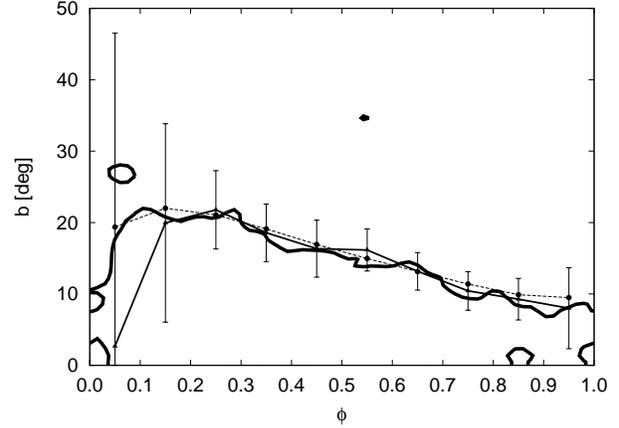}
\caption{Values of $b_{0}=c_{2}/c_{1}$ for 10 phase bins with error bars
(solid line with solid triangles), $\overline{b(\phi)}$ (dashed line with solid circles) and contour
line of $v_{mer}=0$ (thick solid line).}
\label{vmer0_fits}
\end{figure}

Perhaps the most interesting result is that the contour line from Fig.~\ref{mapchart_vmer} representing
values $v_{mer} = 0$ is very close to centre of activity in each phase, i.e. average latitude
as a function of phase, $\overline{b(\phi)}$. Average latitude, $\overline{b(\phi)}$, was calculated
with $\overline{b(\phi)} = \sum b_{i}/n(\phi)$ (sunspot group area was not taken into account).
Therefore, we calculated values of $b_{0}$, for which $v_{mer}=0$,
in all 10 phase bins ($b_{0}=-c_{2}/c_{1}$). The values are given in Table~\ref{linFitCoeffs} next to $\overline{b(\phi)}$.
Both quantities are shown in Fig.~\ref{vmer0_fits}. Average latitudes, $\overline{b(\phi)}$, are drawn with a dashed
line and solid circles, while $b_{0}$ is drawn with a thin solid line, solid triangles and error bars. The error bar
for phase $\phi$=0.05 is off the scale. We also showed the contour line for $v_{mer}=0$ obtained from the map plot
presented in Fig.~\ref{mapchart_vmer}.
Apart from the highly uncertain value of $b_{0}$ for phase $\phi<0.1$ the agreement between $b_{0}$
and $\overline{b(\phi)}$ is very good.

Regions with latitudes closer to solar
equator move toward the pole, while centre of activity in each phase roughly marks the line where
meridional motion changes to equatorward motion.

\subsection{Torsional oscillations}
Torsional oscillations can be described as a pattern in which the solar rotation
is sped up or slowed down in certain regions of latitude. In this section we will investigate
if we can reveal this pattern by examining the $\Delta v_{rot}$ relationship with latitude, $b$,
and cycle phase, $\phi$.

\begin{figure}
\includegraphics[width=84mm]{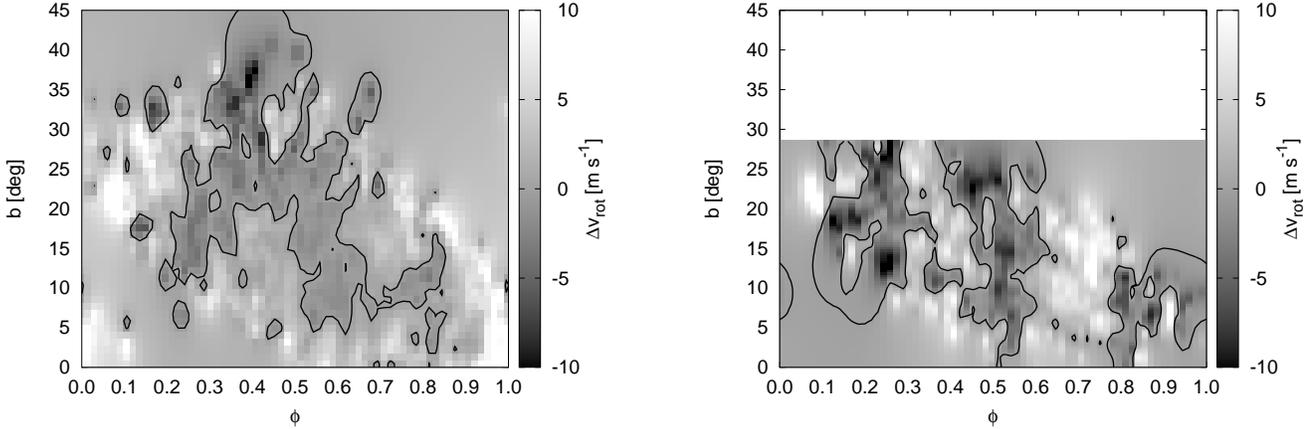}
\caption{Map plot of $\Delta v_{rot}$ as a function of phase of the solar cycle,
$\phi$, and latitude $b$.}
\label{dvrot_map}
\end{figure}
\begin{figure}
\includegraphics[width=84mm]{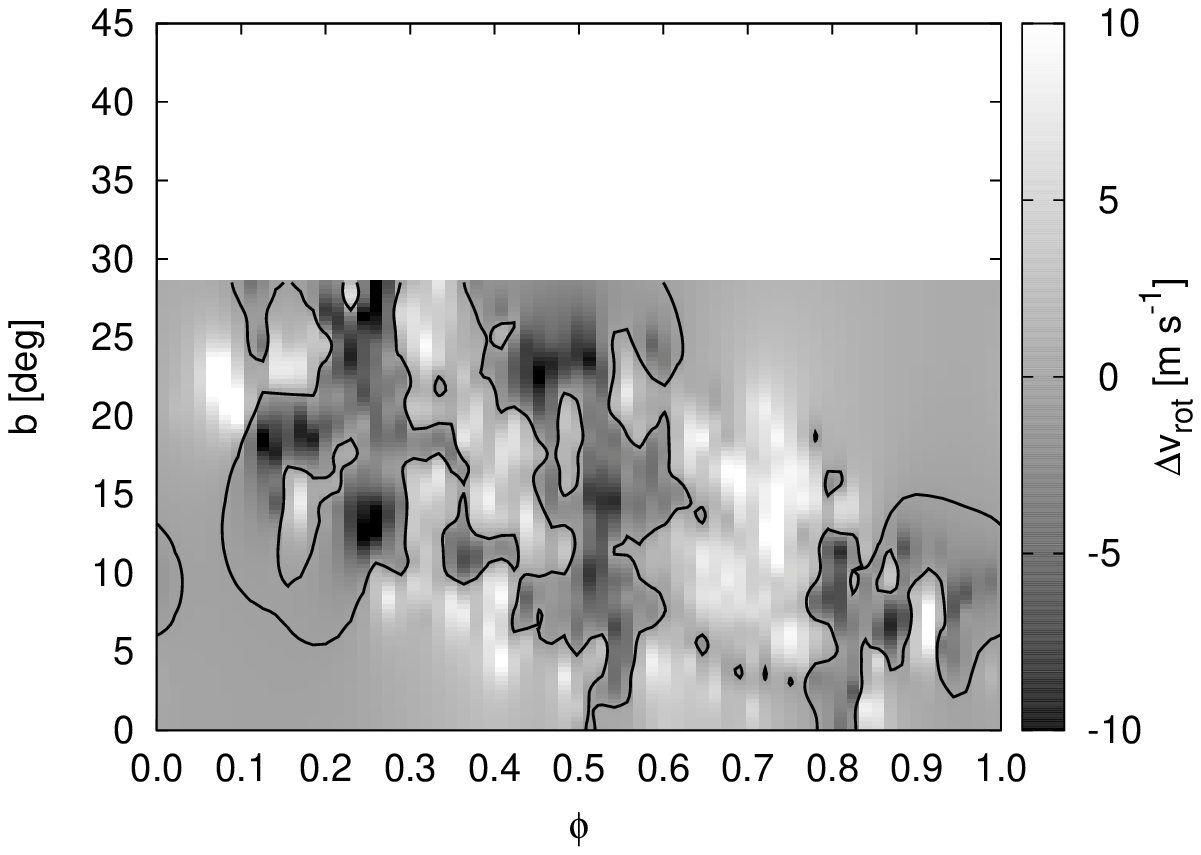}
\includegraphics[width=84mm]{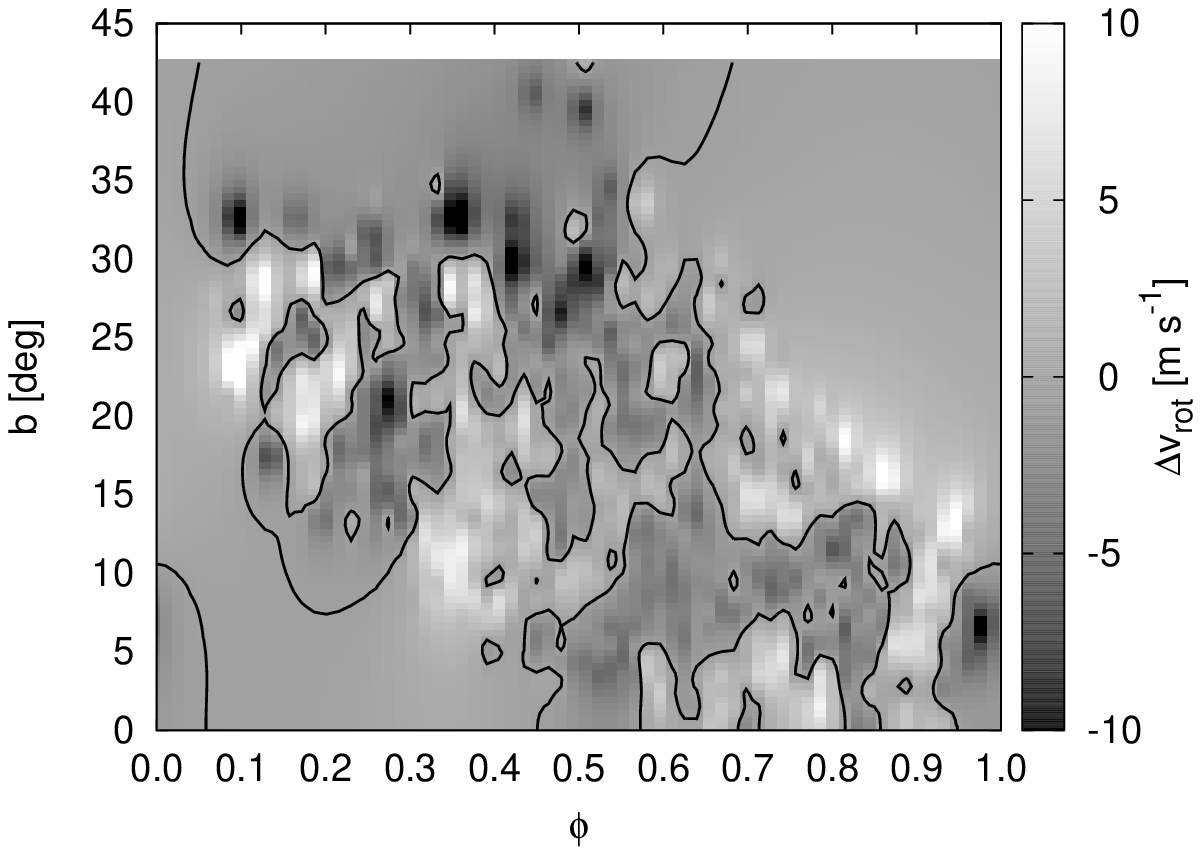}
\includegraphics[width=84mm]{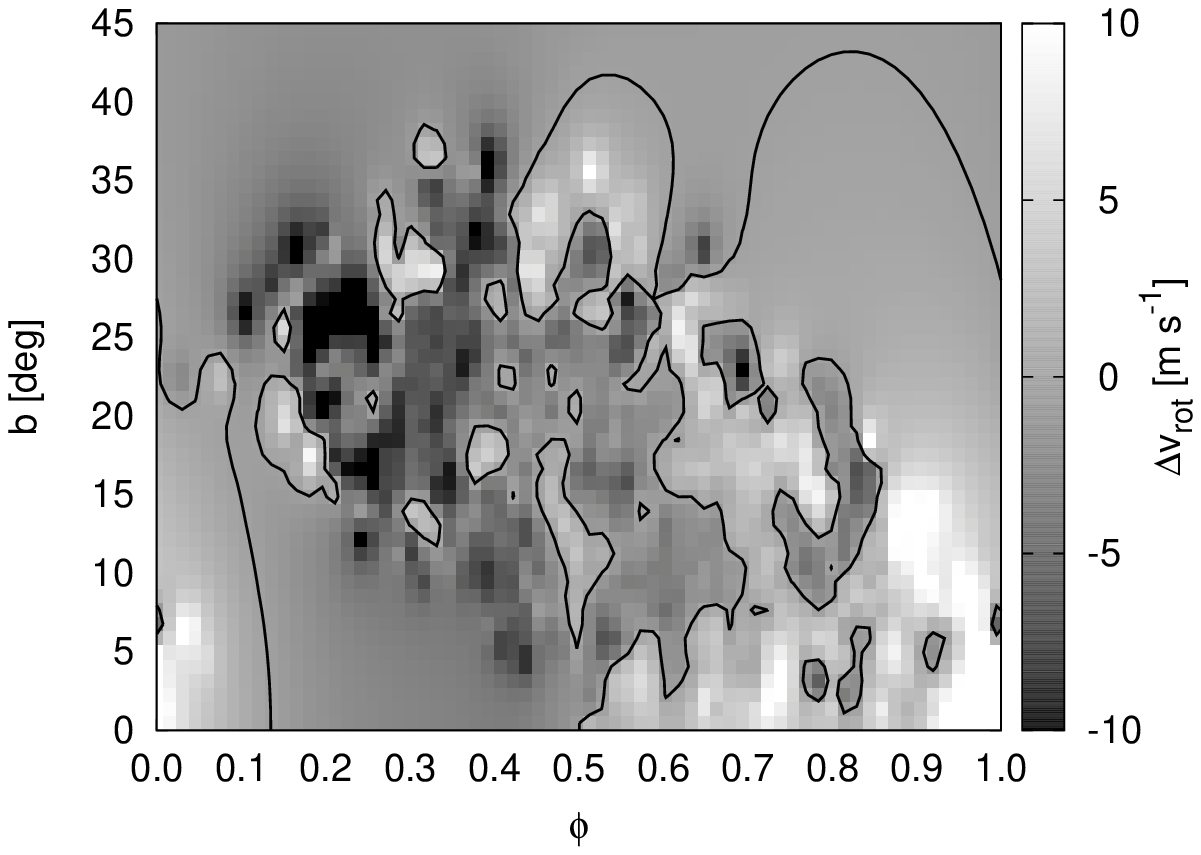}
\caption{The same as in Fig.~\ref{dvrot_map} except that the EGR dataset
is divided into three epochs. From top to bottom we show early epoch (cycles 12 - 15),
mid (cycles 16 - 19) and late (20 - 24). In early epoch (top figure) there is
insufficient number of data points for latitudes above $\approx$30\degr .}
\label{dvrot_map_3epochs}
\end{figure}
We constructed a map plot of $\Delta v_{rot}$ (Fig.~\ref{dvrot_map}) in the same fashion
as in the previous section.
Thick dashed lines show a contour
where $\Delta v_{rot} = 0$, darker regions depict slower than average rotation ($\Delta v_{rot}<0$), while
brighter regions show faster then average rotation ($\Delta v_{rot}>0$).
Fig.~\ref{dvrot_map} shows a rather complex pattern of rotation velocity residuals.
The pattern does not look like a typical torsional oscillation pattern. In order to
investigate if the typical torsional oscillation pattern is only present in today's data,
we divided our dataset into three epochs: early (cycles 12 - 15), mid (cycles 16 - 19) and
late (cycles 20 - 24) and plotted the same
type of plot (Fig.~\ref{dvrot_map_3epochs}). However, the figure doesn't reveal anything
that even resembles the typical torsional oscillation pattern. Moreover, results from
three different epochs are not consistent between themselves and we can't see any
regularity in changes of the pattern over time. With our dataset it is not possible to
make the division into shorter epochs, because as we can see from the top part of
Fig.~\ref{dvrot_map_3epochs} we are already running out of datapoints at latitudes
above 30\degr .

\subsection{Correlation of $v_{mer}$ and $\Delta v_{rot}$ and covariance $<v_{mer}\Delta v_{rot}>$}
In this section we will investigate correlation and covariance of meridional
velocity, $v_{mer}$, and rotation velocity residuals, $\Delta v_{rot}$.
Due to our convention that negative meridional motion reflects equatorward
motion, it follows that negative values of covariance, $q$, means that
angular momentum is also transported toward the solar equator.

\begin{figure}
\includegraphics[width=84mm]{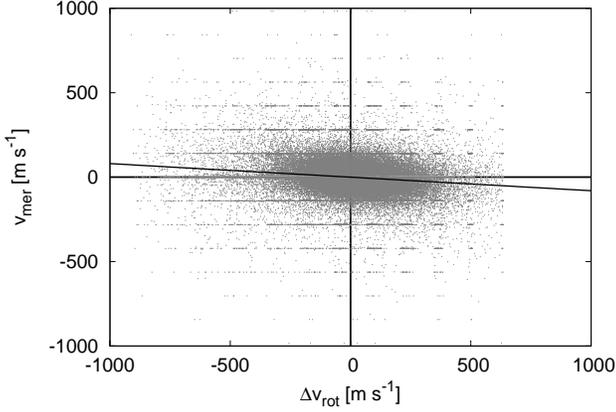}
\caption{Correlation between meridional velocity, $v_{mer}$, and
rotation velocity residual, ${\Delta v_{rot}}$, for
sunspot groups. Linear fit is shown with a solid line.}
\label{IndSpots_dvrot_vmer}
\end{figure}

In Fig.~\ref{IndSpots_dvrot_vmer} we show $v_{mer}(\Delta v_{rot})$ relationship.
The least squares linear fit is described by the
following relation:
\begin{equation}
v_{mer} = (-0.0804\pm 0.0017)\cdot\Delta v_{rot} +(-0.12\pm 0.27)\ \mathrm{m\ s}^{-1}
\label{EqIndSpots_dvrot_vmer}
\end{equation}
and is shown on the figure with the solid line.

\citet{Howard1984} criticised usage of sunspot groups to derive
correlation between $v_{mer}$ and $\Delta v_{rot}$. Because of the average sunspot group tilt
of about 4$\degr$ \citep{Howard1991b} toward the equator one can imagine that as the
group evolves, measured position of the group could be biased toward the
tilt line. Consequently, derived velocities would also be biased and would
produce just the type of correlation between $v_{mer}$ and $\Delta v_{rot}$
as we see in Fig.~\ref{IndSpots_dvrot_vmer} and Eq.~\ref{EqIndSpots_dvrot_vmer}.
However, it is very difficult to actually quantify the magnitude of this effect
and to conclude how much it would influence the true correlation between the
two velocities. By using coronal bright points (CBP) as tracers, \citet{Vrsnak2003}
derived almost the same correlation. The tilt angle is
irrelevant for CBPs and this is our first hint from independent measurements that the
bias mentioned by \citet{Howard1984} is negligible or not present at all.
GPR catalogue also contains the information about the morphological type of the
sunspot group. Quite a significant number of them are actually classified
as a single spot. Tilt angle for such groups is meaningless and no
such bias can exist for this subset. Therefore, we checked the correlation
between $v_{mer}$ and $\Delta v_{rot}$ for this subset of single spots and obtained:
\begin{equation}
v_{mer} = (-0.0803\pm 0.0036)\cdot\Delta v_{rot} +(-1.40\pm 0.46)\ \mathrm{m\ s}^{-1}.
\label{EqIndSpots_dvrot_vmer_zerotype}
\end{equation}
The correlation is virtually identical to the one obtained for all groups
(Eq.~\ref{EqIndSpots_dvrot_vmer}). This is a conclusive proof that the group
tilt bias is not significantly affecting the observed correlation when using
sunspot groups as tracers.
In Fig.~\ref{IndSpots_dvrot_vmer} there are visible artefacts in the form
of horizontal lines which correspond to steps of 1 deg/day. They are a consequence of
poor precision of position in SOON/USAF/NOAA part of the dataset which is usually
recorded with 1 deg precision only. Coupled with usual 1 day period between successive
measurements we get the horizontal artefacts separated by 1 deg/day. However, we note that they do not significantly
affect the results, since whole SOON/USAF/NOAA part of the dataset was excluded from calculations
given in Eq.~\ref{EqIndSpots_dvrot_vmer_zerotype} because this part of the dataset does not
contain the information on the morphological type of the sunspot groups.

\begin{figure}
\includegraphics[width=84mm]{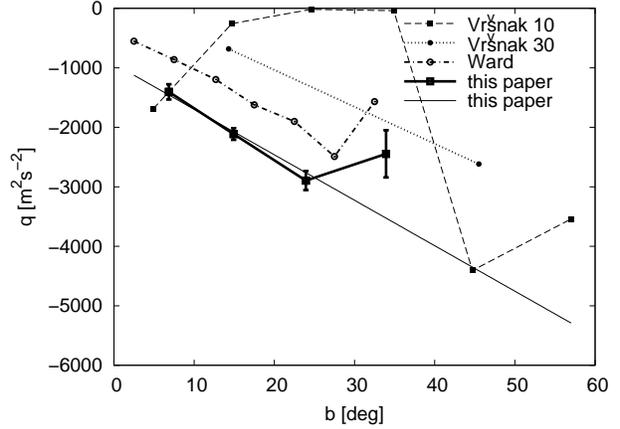}
\caption{Relationship between covariance, $q$, and latitude, $b$, represented
by average values in bins of different width obtained with data from this paper
and other authors. Results from \citet{Ward1965} are labelled Ward and results
from \citet{Vrsnak2003} are labelled Vr\v{s}nak 10 and Vr\v{s}nak 30 for their bins
of 10 and 30 degrees, respectively.
Linear fit (Eq.~\ref{EqIndSpots_q_b}) is shown with a thin solid line.}
\label{average_q_b}
\end{figure}

We also investigated if averages
of $<\Delta v_{rot}v_{mer}>$ grouped in bins of 10\degr\ will
show any dependence of $q$ with latitude. We grouped the dataset into four subsets spanning from
0\degr\ to 10\degr, 10\degr\ to 20\degr, 20\degr\ to 30\degr\ and above 30\degr\ in latitude and
then simply calculated the averages of $\Delta v_{rot}v_{mer}$ product
for each bin to obtain $q(b)$ values. The results
are given in Table~\ref{TabIndSpotsBins} and shown on Fig.~\ref{average_q_b}.
On the same figure we also show
results obtained by \citet{Vrsnak2003} (their 10\degr\ and 30\degr\ bins)
and \citet{Ward1965} and also a linear fit
\begin{equation}
q = (-76.4\pm 9.5)\ \mathrm{m}^{2}\ \mathrm{s}^{-2}\mathrm{(\degr)}^{-1}\cdot b +(-933\pm 161)\ \mathrm{m}^{2}\mathrm{\ s}^{-2}
\label{EqIndSpots_q_b}
\end{equation} through individual data points of our EGR dataset.

\begin{table}
\caption{Average value of covariance, $q$, for several bins in latitude, $b$.}
\label{TabIndSpotsBins}
\begin{center}
\begin{tabular}{c c c c}
\hline
bin & q[m$^2$ s$^{-2}$] & b[\degr] & n\\
\hline
$0^{\circ} < b < 10^{\circ}$&-1404$\pm$128& 6.85 & 35836\\
$10^{\circ} < b < 20^{\circ}$&-2113$\pm$98& 14.93 & 67156\\
$20^{\circ} < b < 30^{\circ}$&-2896$\pm$160& 23.93 & 28518\\
$b > 30^{\circ}$&-2446$\pm$397& 33.953& 4164\\
\hline
\end{tabular}
\end{center}
\end{table}

\begin{figure}
\includegraphics[width=84mm]{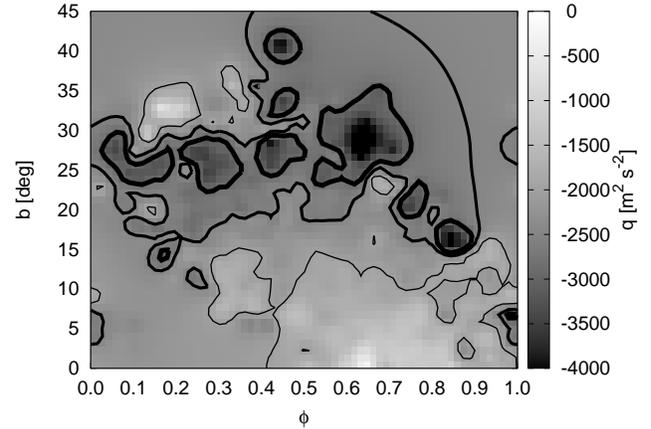}
\caption{Map plot of covariance, $q$, versus phase of the solar cycle, $\phi$, and
latitude, $b$. Levels of $q = -2000$ m$^{2}$ s$^{-1}$, $q = -2500$ m$^{2}$ s$^{-2}$ and $q = -2900$ m$^{2}$ s$^{-2}$
are shown with solid lines, from thinest to thickest, respectively.}
\label{mapchart_q}
\end{figure}

Finally, in Fig.~\ref{mapchart_q} we show a map plot of covariance,
$q$, as a function of phase of the solar cycle, $\phi$, and
latitude, $b$. In the same figure we outlined levels of $q = -2000$ m$^{2}$ s$^{-2}$,
$q = -2500$ m$^{2}$ s$^{-2}$ and $q = -2900$ m$^{2}$ s$^{-2}$
with thinest to thickest solid line, respectively.
This representation of covariance, $q$, is fairly consistent across all phases and latitudes
with larger values of $q$ concentrated around lower latitudes for all phases.

\section{Discussion}
Before we comment on the results presented, we must clarify statistical significance of various
types of plots shown in this work. Map plots have the lowest statistical significance because they
show dependence of the variable with respect to two parameters which means that in each bin
there are significantly fewer data points than in other plots. In this work these plots were
used only as informative and not a single value was calculated from them. They do, however
give a useful first look at data and various relationships.

Second type of plots show some empirical curve fitted to the data with one variable and one parameter.
These have higher importance because density of data points is significantly larger in one-dimensional
parameter space than in two-dimensional. Nevertheless, caution must be taken before interpretation. Since fitted curves
are only empirical, the form of the equation largely depends on how the raw data look like, rather
then what it should be like from theoretical modelling. Typical example is shown in Fig.~\ref{vmer_b_aver}
where the linear function is fitted to the data. The look of the function might lead someone to
believe that there is a positive meridional flow at 0 deg latitude. However, if we look at average
values in bins of latitude (even more conclusively shown in the inset to Fig.~\ref{vmer_b_aver}), we
see that the flow actually drops to zero for very low latitudes on both hemispheres. Values of parameters of the
fitted linear function are dominated by mid-latitude behaviour where most of the data points actually are.

The third type of plots are plots where we show average values grouped in bins. Together with their errors,
they should be the most reliable representation of what true relationship actually looks like.

\subsection{Meridional motions}
We have seen that meridional motions of sunspot groups show a distinct pattern where
at latitudes below the centre of activity, the motion is poleward, while on the other side the
motion is predominantly toward the equator.
In addition we have shown that this is valid for all phases of the solar cycle.
Flow converging to activity belts from both sides suggests that the plasma is
circulating towards the centre of the Sun at these latitudes (sinking). Diverging flow on the
equator suggest that the plasma is circulating from below to the surface.
There might be another latitude point of diverging flow at about 40\degr , but
our results are inconclusive if this diverging flow is real.

This result is in contradiction with most of the other results generated from tracer measurements
\citep{Howard1991a,Howard1996,Snodgrass1996,Vrsnak2003}. In most of them the flow is opposite
to ours; they found meridional flow to be out from the centre of activity. Although, none
of the above papers mention to which latitude calculated velocity is assigned to, we suggest that
their result is a consequence of not properly accounting for non-uniform latitudinal distribution of tracers they used.
Moreover, if we use any of the other two possibilities (average latitude or latitude of the second measurement of position)
we get almost exactly the opposite flow.
Our approach, used in this paper, eliminates the need to keep track of tracer distribution and shows the true
meridional motion.

Helioseismology measurements usually detect predominant poleward flow at all latitudes
\citep{Zhao2004, Gonzalez2008, Gonzalez2010}. However, there is a striking similarity between
our results for meridional flow (Figs.~\ref{mapchart_vmer} and \ref{vmer_b_aver}) and {\em residual}
meridional flow found by helioseismology \citep{Zhao2004, Gonzalez2008, Gonzalez2010}.
For example all details from our Fig.~\ref{vmer_b_aver} are reproduced in \citet{Gonzalez2008}.
This includes the change from equatorward to poleward flow at about 40\degr of latitude, even though
our results are highly uncertain at these high latitudes.
\cite{Zhao2004} explicitly state that residual meridional flow is converging toward activity
belts.
This raises the question, why we can't see the dominant poleward flow in tracer measurements?
One possibility is that this flow is not constant in time and that it was different in the past
which averages to net zero flow and we only observe residual which is permanent.

\citet{Hathaway2012} used supergranules as probes of the Sun's meridional circulation.
He found that surface poleward flow gradually changes to equatorward as we go deeper
below solar surface. One of the intermediate between those two extremes might be the
results we obtained by tracing sunspot groups. \citet{Ruzdjak2004}
suggested that at their birth sunspots groups could be coupled to the layer
at about $r=0.93$ $R_{\odot}$ effectively showing the plasma flow from beneath
the solar surface.

Another possibility to explain the difference between helioseismology and tracer
measurements is that the flow is different around sunspots than in the rest
of solar surface. With tracers
we are confined to a small region of the solar disk around active regions, while
helioseismology does not have this limitation.

\subsection{Torsional oscillations}
Torsional oscillation pattern is usually described as distinct bands of faster
and slower than average rotation rate. These bands move toward the equator with time at low latitudes
\citep[see for example][]{Basu2003}. 

Our analysis of rotation velocity residuals reveals a pattern much less distinctive than
for the meridional flow. Apart from generally remarking that slower than average flow
is obtained around the maximum and faster than average around the minimum there is hardly
anything else we could say about it. Similar behaviour was found by \citet{Brajsa2006}
and \citet{Brajsa2007}.
It has no distinct latitudinal dependence and it shows no significant correlation with
zones of activity.
\citet{Zhao2004} showed zonal flows for years 1996-2002 (solar cycle 23) and it looks as if their
results are similar to ours (faster than average in the phases around the minimum of solar
activity and slower than average around the maximum).
However, we can't confirm that the typical torsional oscillation pattern is visible
in sunspot data.

We investigated the rotation residual flow in three different epochs to see if the
typical torsional oscillation pattern can only be seen in modern data. The pattern does
change with time, but it has no resemblance to torsional oscillations. Moreover, we
can't see any regularity in it's shape and change from epoch to epoch. Thus we were
unable to quantify dependence of rotation residuals with respect to phase and latitude.
It's quite possible that we see only random noise.

\subsection{Reynolds stress}

From our Fig.~\ref{average_q_b}, Eq.~\ref{EqIndSpots_q_b} and Table~\ref{TabIndSpotsBins}
it is easily seen that the equatorward angular momentum transfer
is predominant at all latitudes covered by sunspot group data ($q<0$). This is consistent with other
studies using different methods and/or different data samples \citep{Ward1965, Gilman1984, Vrsnak2003}.

Average values of $q$ in 10\degr\ bins of $b$, shown in Table~\ref{TabIndSpotsBins}, shows very similar
behaviour to the linear fit for sunspot groups sample. \citet{Vrsnak2003} showed similar results for 10\degr\ bins using CBP
sample (their Fig.~6c), but the qualitative behaviour is different than ours. We suggest that this
is due to their 10\degr\ bins being of too low statistical significance. This is also implied by their
30\degr\ bins which show different qualitative behaviour much more similar to ours.
Moreover, our results for 10\degr\ bins show very similar behaviour to that of \citet{Ward1965}, both qualitatively
and quantitatively.

\begin{figure}
\includegraphics[width=84mm]{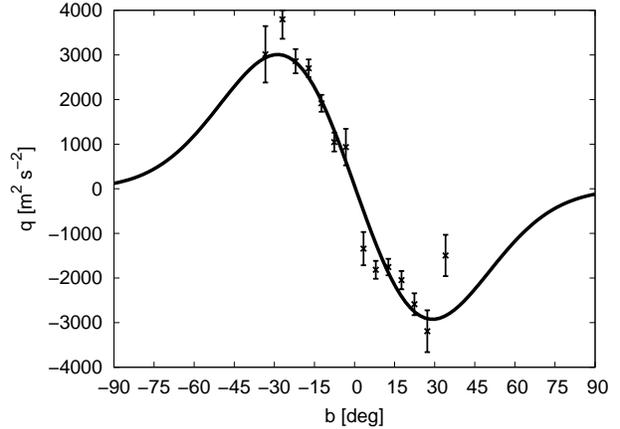}
\caption{$q(b)$ relationship for EGR sample fitted with exponential cutoff 
model (solid line) and average values of $q$ in bins spanning both solar
hemispheres.}
\label{fits}
\end{figure}

Another interesting feature visible in Fig.~\ref{average_q_b} is that
there appears to be a minimum in $q(b)$ relationship around 25-30\degr, both for our sample and
for that given by \citet{Ward1965}. This is also reminiscent of similar
result obtained by, for example, numerical simulations \citep{Pulkkinen1993} and results obtained by \citet{Canuto1994}
based on theoretical considerations.
In order to investigate this a little more we plotted average values of covariance, $q$, with respect to latitude
for both solar hemispheres separately (Fig.~\ref{fits}). In this representation negative values represent
angular momentum transfer toward equator for northern solar hemisphere. On the southern hemisphere
{\em positive} values of $q$ also show momentum transfer toward the equator. So, this Fig. is consistent with
Figs.~\ref{average_q_b} and \ref{mapchart_q} where we see equatorward momentum transfer at all
available latitudes.

Since the minimum at around $b=$25-30\degr of latitude appears at all our plots we constructed an
empirical relationship between $q$ and $b$ which takes into account dominant linear dependence
for low latitudes and allows for a possible minimum at some unspecified latitude:
\begin{equation}
\label{EqModelE}
q = (e_{1}b + e_{2})\cdot e^{-e_{3}b^{2}}.
\end{equation}
This shape can produce a minimum, but could also 'explode' to $q=\pm\infty$, if $e_{3}$ turns
out to be negative.
By fitting through individual data points we obtained the coefficients of the fit shown in
Table~\ref{TabNonLinearFits} and we also show the fit in Fig.~\ref{fits} as a solid line.

The shape of the curve is very similar to that given by \citet{Canuto1994} in their
Fig.~15 including the prediction that $q(b)$ falls to zero at the poles. The form
of our fit function (Eq.~\ref{EqModelE}) does not guarantee that, because
the answer to the question if
and where (in terms of $b$) covariance, $q$, falls to zero
is highly sensitive to fitted coefficients and not constrained by the physical fact
that maximum latitude is $b=$90\degr.

However, we must stress that although the agreement with \citet{Canuto1994} is good
and that calculated coefficients seem strongly constrained (Table~\ref{TabNonLinearFits}),
latitudinal extent of sunspot groups is very limited and anything that happens
beyond $b=$35-40\degr is ambiguous considering our dataset.
Even the location and depth of the minimum is not very reliable for
the same reason.
Therefore, it is of great importance to confirm or disprove our hypothesis
about the shape of $q(b)$ at mid to high latitudes.

The depth of the minimum, is also close to the one found by \citet{Canuto1994} considering
that $q\approx -3000$ m$^{2}$ s$^{-2} \approx$ -0.15(\degr day$^{-1}$)$^{2}$.

\begin{table}
\caption{Non linear fits coefficients for exponential cutoff model.}
\label{TabNonLinearFits}
\begin{center}
\begin{tabular}{c c}
\hline
coefficient & value\\
\hline
$e_{1}$ [m (\degr)$^{-1}$ s$^{-2}$] & $-169.0\pm9.7$\\
$e_{2}$ [m$^{2}$ s$^{-2}$] & $69\pm80$\\
$e_{3}$ [(\degr)$^{-2}$] & $0.00060\pm0.00012$\\
\hline
\end{tabular}
\end{center}
\end{table}

\section{Summary and Conclusion}
The most important results can be summarised as follows:
\begin{itemize}
\item{Meridional motion of sunspot groups clearly shows poleward motion for all latitudes
bellow the centre of activity and motion toward the equator for higher latitudes. This
is valid for all phases of the solar cycle with a very strong correlation.}
\item{The variations of $v_{mer}$ with latitude, $b$, are approximately in
the range of $v_{mer}=\pm 10$ m s$^{-1}$.}
\item{Rotation velocity residuals show unusual torsional oscillation pattern. The actual values of rotation residual velocity
rarely exceed $\Delta v_{rot}=$5 m s$^{-1}$. Rotation velocity residuals map plots in different epochs show
a changing pattern which we are unable to explain. It is possible that we see only a pattern
resulting from random errors.}
\item{Meridional velocities are similar to {\em residual} meridional velocities found
with time-distance helioseismology \citep{Zhao2004, Gonzalez2010}.}
\item{Reynolds stress is negative at all available latitudes which, in our convention,
corresponds to equatorward transport of angular momentum. This supports the idea that
observed rotational profile is actually driven by the Reynolds stress.}
\item{Latitudinal dependence of Reynolds stress suggests a minimum at about $b\approx 30^{\circ}$.
The value of covariance being $q\approx$-0.15 (\degr)$^{2}$ day$^{-2}$ ($q\approx$ - 3000 m$^2$ s$^{-2}$).
This is consistent with \citet{Canuto1994}.}
\item{Phase and latitudinal dependence of covariance, $q$, seems to be fairly uniform in all
phases of the solar cycle with possible exception at phases very late in the solar cycle.}
\end{itemize}

Most authors who used tracers to track the meridional flow found meridional velocities
to be opposite to ours. We believe that this is a consequence of improper assignment of
latitude to measured velocities. In the approach we used (assigning velocities to the
starting latitude), latitudinal distribution
of tracers is irrelevant because calculation of averages (and even fit functions)
does not need to take into account the number of tracers at specific latitude ($n(b)$ is
the same for all of them for each particular $b$). If we were to use average
latitude between two successive measurement of position, we would have to calculate
{\em weighted} averages by taking into account from which latitude the tracer
actually started and assign weight accordingly.

As a test, we have also made an analysis with assigning the second latitude of two successive
measurements and calculated average meridional motions and got the results very similar
to, for example, \citet{Snodgrass1996} and \citet{Vrsnak2003}. Assigning second latitude or
average latitude suffers from exactly the same problem; we would need to take into account
the first latitude in order to properly calculate
weighted averages. By using the starting latitude as the relevant one, we simply
defined the flow as flow {\em from} certain latitude instead of flow {\em into}
some latitude. There is no loss of generality in doing so and no difference
in physical interpretation.
Similar result for meridional velocity to ours was obtained by \citet{Olemskoy2005} who
pointed out the solution to this problem and used starting latitudes of the tracers.
As a point of interest, even when we used the second latitude in our test, calculated
correlation and covariance of meridional and residual rotation velocities was very similar
to the results we obtained in this paper by using the first latitude. This also explains why our results,
regarding correlation and covariance, are similar to the results of other authors who
used tracers even if they found different average meridional flow than we did.

We can see a clear increase of uncertainties at latitudes larger than 30\degr\ which
is a consequence of sunspot latitudinal distribution. Therefore, it is of great importance
to use other methods or tracers (for example CBPs) to extend the analysis
to higher latitudes.

The absence in our data of predominant
poleward meridional flow which is found in helioseismology might be explained by
several possibilities. The first one is that sunspots are anchored at depth below
the surface showing sub-photospehric flow. Another possibility is that on longer
time-scales the flow changes from poleward to equatorward and consequently averages
out in our analysis. And finally, it is possible that the flow in active regions
is different than in the rest of the solar disk, so in our analysis we see only
this localised flow. Future work might shed some light on these opened questions.

\section*{acknowledgements}
We acknowledge the work of all the people that contribute to and maintain the GPR and SOON/USAF/NOAA Sunspot
databases. We also thank the anonymous referee for helping to significantly improve the quality
of this paper. The research was partly funded from
the European Commission's Seventh Framework Programme (FP7/2007-2013) under the grant
agreements n\degr 284461 [eHEROES] and n\degr 312495 [SOLARNET].

\bibliographystyle{mn2e}
\bibliography{sunQ}

\end{document}